\documentclass[aps,pre,notitlepage,nofootinbib]{revtex4-2}

\usepackage{graphicx}
\usepackage{amsmath,amssymb}

\newcommand{\cG}{{\cal G}}
\newcommand{\eg}{{e.g., }}
\newcommand{\ie}{{i.e., }}
\newcommand{\tcG}{\tilde{\cal G}}
\newcommand{\tp}{\tilde{p}}
\newcommand{\tvecu}{\tilde{\bf u}}
\newcommand{\tvecv}{\tilde{\bf v}}

\newcommand{\vecq}{{\bf q}}
\newcommand{\vecr}{{\bf r}}
\newcommand{\vecrho}{\boldsymbol{\rho}}
\newcommand{\vecu}{{\bf u}}
\newcommand{\vecv}{{\bf v}}
\newcommand{\khat}{\hat{k}}

\begin{document}

\title{Structured viscoelastic substrates as linear foundations}

\author{Chen Bar-Haim}

\author{Haim Diamant}

\email{hdiamant@tau.ac.il}
%\url{http://www.tau.ac.il/~hdiamant}

\affiliation{Raymond and Beverly Sackler School of Chemistry, and Center for the Physics and Chemistry of Living Systems, Tel Aviv University, Tel Aviv 6997801, Israel}

\begin{abstract}

The linear (Winkler) foundation is a simple model widely used for decades to account for the surface response of elastic bodies. It models the response as purely local, linear, and perpendicular to the surface. We extend this model to the case where the foundation is made of a structured material such as a polymer network, which has characteristic scales of length and time. We use the two-fluid model of viscoelastic structured materials to treat a film of finite thickness, supported on a rigid solid and subjected to a concentrated normal force at its free surface. We obtain the foundation modulus (Winkler constant) as a function of the film's thickness, intrinsic correlation length, and viscoelastic moduli, for three choices of boundary conditions. The results can be used to readily extend earlier applications of the Winkler model to more complex, microstructured substrates. They also provide a way to extract the intrinsic properties of such complex materials from mechanical surface measurements.

\end{abstract}

\maketitle

\section{Introduction}
\label{sec_intro}

The Winkler (or linear) foundation is arguably the simplest model for the surface response of an elastic solid \cite{JohnsonBookChapter4,MansfieldBook,Dillard2018}. It assumes that a normal surface displacement $u_z(\vecrho)$ at a point $\vecrho=(x,y)$ on the surface, and the restoring force per unit area, $f_z(\vecrho)$, are related  linearly and completely locally,
\begin{equation}
    f_z(\vecrho) = -k u_z(\vecrho),
\label{Winkler0}
\end{equation}
as if a local perpendicular spring of spring constant (per unit area) $k$ resisted the displacement. Since its introduction over a century and a half ago, the Winkler foundation has been used to model a remarkable variety of surface phenomena. For a recent review, see Ref.~\cite{Dillard2018}. These include, in particular, the adhesion, delamination, and buckling of slender bodies (beams, plates, shells) supported on elastic or liquid substrates. Recent representative works are on the localized buckling of supported thin sheets \cite{Audoly2011}, and the delamination of sheets and cylindrical shells off soft adhesive substrates \cite{Oshri2018,Oshri2020}. Various extensions and refinements of the model have been developed over the years \cite{Dillard2018}.

As noted already by Biot \cite{Biot1937}, there are questions concerning the relation of Eq.~(\ref{Winkler0}) to the properties of the elastic substrate. For example, the Boussinesq problem \cite{LLelasticityChapter1,JohnsonBookChapter3}, addressing the response of a semi-infinite elastic bulk to a concentrated surface force, yields a highly non-local response, with the normal displacement decaying only as $1/\rho$ from the point of forcing. This is a consequence of the absence of length scale in linear elasticity. Specifically, taking $f_z(\vecrho)=F_z \delta(\vecrho)$, we have \cite{LLelasticityChapter1}
\begin{equation}
     u_z(\vecrho') = \frac{1-\nu}{2\pi G}\,\frac{1}{|\vecrho-\vecrho'|}
    \, F_z,
\label{Boussinesq0}
\end{equation}
where $G$ is the material's shear modulus and $\nu$ its Poisson ratio. The contrast between the two pictures is seen also in the difference between the units of $k$ (force/length$^3$) and those of the substrate's elastic moduli (force/length$^2$). The missing length scale may be provided by the wavelength of an imposed undulating displacement \cite{Biot1937,Brau2011}, or by a finite thickness $h$ of the substrate \cite{JohnsonBookChapter4,Vlasov1966}. In the latter, more common case, assuming that the substrate is attached on its other side to a much more rigid solid, the surface response is cut off at distances $\rho\gtrsim h$ from the forcing point. Thus, if one examines the surface over length scales significantly larger than $h$, the response will look sharply localized and may be approximated by a delta function, reproducing Eq.~(\ref{Winkler0}) with $k\sim G/h$. Specifically, the following relation has been derived \cite{Vlasov1966}:
\begin{equation}
    k = \frac{2(1-\nu) G}{(1-2\nu)h}.
\label{Vlasov}
\end{equation}

Formal asymptotic analysis \cite{Baldelli2015} and systematic expansion \cite{Chandler2020} have been presented, obtaining the Winkler foundation as the small-$h$ reduction of three-dimensional elasticity. In the present work we introduce another scheme of systematic reduction. The reductions turn a flat elastic surface effectively into a layer of identical perpendicular springs of fixed lateral density. If the material is incompressible, the change in density caused by loading the perpendicular springs must be compensated by a change in their lateral density (\ie change in $k$), which the model does not allow. This is the origin of the divergence of $k$ as given by Eq.~(\ref{Vlasov}) for $\nu\rightarrow 1/2$. Thus the incompressible limit requires a different theory \cite{Dillard1989,Dillard2018,Chandler2020}. A systematic expansion in small $h$, which has recently been presented \cite{Chandler2020}, obtains the response to a surface force as a combination of a term proportional to the local force and another term proportional to its surface second derivative. The additional term removes the divergence in the incompressible limit.

The most straightforward way to generalize the results above to the case of a linear viscoelastic foundation \cite{JohnsonBookChapter6} is to consider a frequency-dependent response, $f_z(\vecrho,\omega)=-k(\omega) u_z(\vecrho,\omega)$,
\begin{equation}
    \label{eq_vlasov}
    k(\omega) = \frac{2(1-\nu(\omega)) G(\omega)}{(1-2\nu(\omega))h},
\end{equation}
resulting from the foundation's complex viscoelastic moduli. In the present work we would like to go further. Most soft materials have at least one intrinsic length scale $\xi$ which is much larger than the molecular size \cite{WittenBook,DoiBook}. For example, in polymer gels $\xi$ is the so-called mesh size\,---\,the characteristic distance between cross-links or entanglement points in the polymer network. This correlation length is typically a few nanometers to a few tens of nanometers, and in biological gels such as semi-flexible actin networks it may reach a fraction of a micron. The intrinsic length scales are accompanied by intrinsic time scales, making the viscoelastic response of these materials both time- and space-dependent \cite{DoiEdwardsBook,GrosbergJPCB2016}. 

Our goal is to calculate the generalized coefficient $k(\omega)$ for such a viscoelastic structured substrate. The potential benefit is twofold. First, predictions based on the elastic Winkler foundation concerning various phenomena such as wrinkling and delamination may be extended to a wider range of complex substrates. Conversely, this variety of phenomena (for example, wrinkling of supported sheets) may be exploited to tap into the structure and viscoelasticity of the supporting material.
Our continuum approach is valid for length scales larger than the correlation length (mesh size). Thus the relevant regime for our analysis of the Winkler-like limit is
\begin{equation}
    \xi \ll h \ll \lambda,
\end{equation}
where $\lambda$ is the length scale of experimental interest (\eg the wrinkle wavelength).

Recently we have presented a solution to the Boussinesq problem for structured materials, \ie the surface response of a semi-infinite bulk of such a material \cite{BarHaim2020}. In the present work we analyze the opposite limit, of a thin film attached to a rigid solid \cite{Kumari2020}. As in Ref.~\cite{BarHaim2020}, the analysis is based on the two-fluid model\,---\,a minimal model of a structured material, made of viscous and (visco)elastic coupled components, with emergent length and time scales \cite{DeGennes1976,DeGennes1976II,DoiOnuki1992,Milner1993,DiamantEpje2015}.

In Sec.~\ref{sec_model} we briefly review the two-fluid model and define its specific application for the present problem. Section~\ref{sec_results} presents the results, focusing on the effective Winkler coefficient $k$ and its dependence on the correlation length $\xi$ and the viscoelastic modulus of the film. In Sec.~\ref{sec_discussion} we discuss implications of the results for distinctive behaviors which can be checked experimentally, and  their possible use for the characterization of structured materials.

\section{Model}
\label{sec_model}

\begin{figure}
 % \centerline{\includegraphics[trim={5cm 0 0 0},clip]{SchemeFiniteFilm.pdf}}
  \centerline{\includegraphics[width=\textwidth,trim={2cm 8cm 0 2cm},clip]{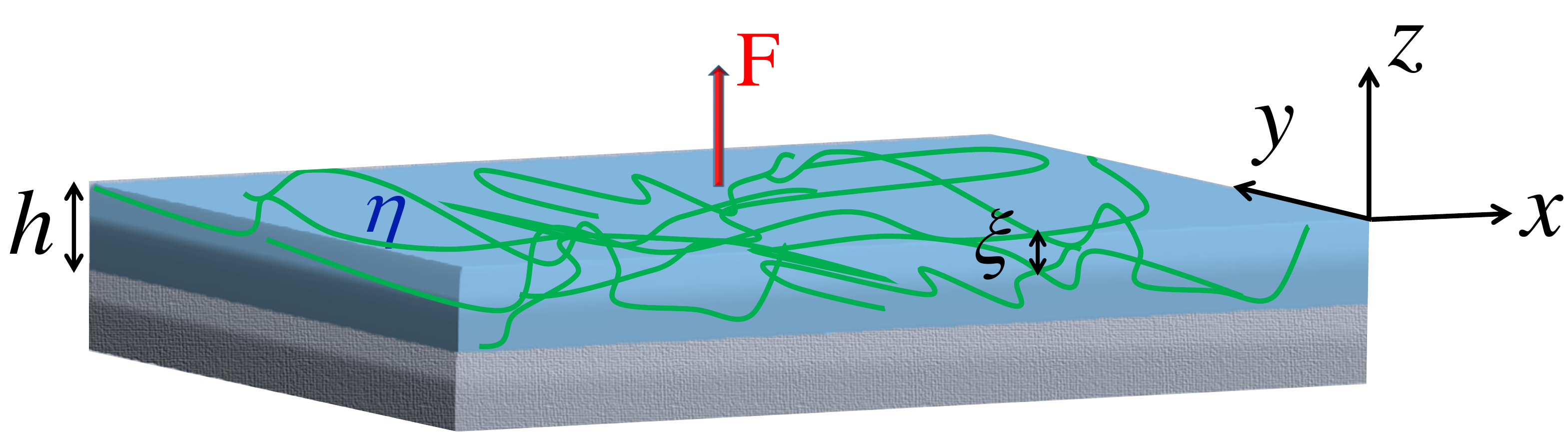}}
\caption{Schematic view of the system and its parameters.}
\label{fig_scheme}
\end{figure}

We consider a film made of a structured medium and occupying the region
$-h<z<0$. At the $(x,y,z=-h)$ plane the film is attached to a rigid substrate. At the $(x,y,z=0)$ plane it is free apart from a localized perpendicular force applied at the origin, $f_z(\vecrho)=F_z\delta(\vecrho)$, where $\vecrho=(x,y)$. See Fig.~\ref{fig_scheme}. 

We use the two-fluid model \cite{DeGennes1976,DeGennes1976II,DoiOnuki1992,Milner1993,Levine2Fluid2001,DiamantEpje2015}
to describe the structured medium.
The model has two components\,---\,a semi-dilute polymer network and a
structureless host fluid. See the schematic illustration in
Fig.~\ref{fig_scheme}. The network is described as a (visco)elastic
medium, whose deformation is defined by a displacement field
$\vecu\left(\vecr,\omega\right)$, which is a function of position
$\vecr=(\vecrho,z)$ and frequency $\omega$. The
corresponding stress tensor is
\begin{subequations}
\begin{equation}
\sigma_{ij}^{(u)}=2G_0\left[u_{ij}-\left(u_{kk}/3\right)\delta_{ij}\right]+K_0u_{kk}\delta_{ij},
\label{StressTensorElasticNet}
\end{equation}
where $u_{ij}\equiv\left(\partial_iu_j+\partial_ju_i\right)/2$ is the
network's strain tensor, and $G_0$ and $K_0$ its shear and compression
moduli, which may be frequency-dependent. In what follows we use instead of $K_0$ the network's Poisson ratio, $\nu_0=(3K_0-2G_0)/[2(3K_0+G_0)]$. The host fluid is described as
viscous and incompressible, having a flow velocity field
$\vecv\left(\vecr,\omega\right)$, pressure field
$p\left(\vecr,\omega\right)$, and the stress tensor
\begin{equation}
\sigma_{ij}^{(v)}=-p\delta_{ij}+2\eta_0 v_{ij},
\end{equation}
\label{StressTensors}
\end{subequations}
where $v_{ij}\equiv\left(\partial_iv_j+\partial_jv_i\right)/2$ is the
fluid's strain-rate tensor, and $\eta_0$ its shear viscosity.

The two components are coupled through mutual friction characterized
by a coefficient $\Gamma$. This coefficient 
is related to the correlation length $\xi$,
$\Gamma\sim\eta_0/\xi^2$ \cite{HardenPincus1st}; see also below. The frictional force density is proportional to the local relative velocity of the two components, $\vecv-i\omega\vecu$,
thus maintaining translational symmetry. Neglecting inertia, we write
the governing equations for the three fields
$(\vecu,\vecv,p)$ as
\begin{subequations}
\begin{eqnarray}
&&0=\nabla\cdot\sigma_{ij}^{(u)}-\Gamma\left(i\omega\vecu-\vecv\right),\label{eq_motion_net}\\
&&0=\nabla\cdot\sigma_{ij}^{(v)}-\Gamma\left(\vecv-i\omega\vecu\right),\label{eq_motion_fluid}\\
&&0=\nabla\cdot\vecv. \label{eq_incomp}
\end{eqnarray}
\label{eqs_motion}
\end{subequations}
The first two equations describe the local balance of forces on the two components. The third accounts for the incompressibility of the host fluid.\footnote{In principle, the incompressibility constraint should apply for the whole composite material; yet, assuming a semi-dilute network, one may apply it approximately to the fluid component alone.}

Equations~(\ref{StressTensors}) and (\ref{eqs_motion}) are supplemented by boundary conditions at the two bounding planes. At the rigid substrate both components are taken to be stationary,
\begin{equation}
  \vecu(\vecrho,-h)=
  \vecv(\vecrho,-h)=
  \nabla_{\vecrho} \,p(\vecrho,-h)=0.
\label{bc_rigid_substrate}
\end{equation} 
At the free surface we consider three cases corresponding to three different experimental limits:
\begin{enumerate}
\begin{subequations}
\label{BCs}

    \item[{\bf BC1}:] The force is applied to both components, and they are strongly coupled,
\begin{equation}
  \sigma_{iz}^{(u)}\left(\vecrho,0\right)+\sigma_{iz}^{(v)}\left(\vecrho,0\right)= f_z(\vecrho)\delta_{iz},\ \ \ 
  i\omega\vecu(\vecrho,0)=\vecv(\vecrho,0).
  \label{bc1}
\end{equation}

    \item[{\bf BC2}:] The force is applied to the network only, and the two components are strongly coupled,
\begin{equation}
  \sigma_{iz}^{(u)}\left(\vecrho,0\right)= f_z(\vecrho)\delta_{iz},\ \ \ 
  i\omega\vecu(\vecrho,0)=\vecv(\vecrho,0).
  \label{bc2}
\end{equation}
    
    \item[{\bf BC3}:] The force is applied to the network only, and the two components are weakly coupled (\ie the fluid surface is left stress-free),
\begin{equation}
  \sigma_{iz}^{(u)}\left(\vecrho,0\right)= f_z(\vecrho)\delta_{iz},\ \ \ 
  \sigma_{iz}^{(v)}\left(\vecrho,0\right)=0.
  \label{bc3}
\end{equation}

\end{subequations}    
\end{enumerate}
Equations
(\ref{StressTensors})--(\ref{BCs}) define our
model.

The two-fluid model has a characteristic time scale, $\eta_0/G_0$, and a characteristic length scale, $(\eta_0/\Gamma)^{1/2}$. Based on these we define, as in earlier works, 
\begin{equation}
  G \equiv G_0 + i\omega\eta_0,\ \ \ 
  \xi\equiv\left(\frac{G_0\eta_0}{G\Gamma}\right)^{1/2},
\end{equation}
as the viscoelastic shear modulus and correlation length of the composite material.
In the limit of low frequency we have $G\simeq G_0$, and $\xi\simeq(\eta_0/\Gamma)^{1/2}$ as anticipated above. The limit of a structureless medium is obtained for $\xi\rightarrow 0$, \ie $\Gamma\rightarrow\infty$, whereby the two components are strongly coupled everywhere and move as one.

\section{Results}
\label{sec_results}

Solving the linear model, Eqs.~(\ref{StressTensors})--(\ref{BCs}), is  straightforward but cumbersome, producing complicated expressions. We  outline the steps of solution,\footnote{In the present special case of a normal surface force, the solution can be simplified by exploiting the problem's axial symmetry. For the sake of future extensions of the theory, we describe the more general derivation.} which can be simply reproduced, and then focus on the main results and useful asymptotic limits.

The first step is to transform all the functions into Fourier space along the two lateral axes, 
$\tilde{g}(\vecq,z,\omega) \equiv \int d\vecrho e^{-i\vecq\cdot\vecrho} g(\vecrho,z,\omega)$. This transforms Eqs.~(\ref{eqs_motion}) into a set of seven second-order ordinary differential equations in $z$ for $\tvecu(\vecq,z,\omega)$, $\tvecv(\vecq,z,\omega)$, and $\tp(\vecq,z,\omega)$. The concentrated surface force density is transformed into a constant, $\tilde{f}_z(\vecq)=F_z$.

The second step is to obtain the general solution of these equations,  which is simplified by decoupling the fields $\vecu$ and $\vecv$ \cite{DiamantEpje2015}. The solution contains 14 unknown amplitudes. They are found by imposing the 14 boundary conditions (\ref{bc_rigid_substrate})--(\ref{BCs}). This provides the solution to all fields.

\begin{figure}[ht]
%\centering{\includegraphics[width=0.7\linewidth]{GpRound3}}
\centerline{\includegraphics[width=0.6\linewidth]{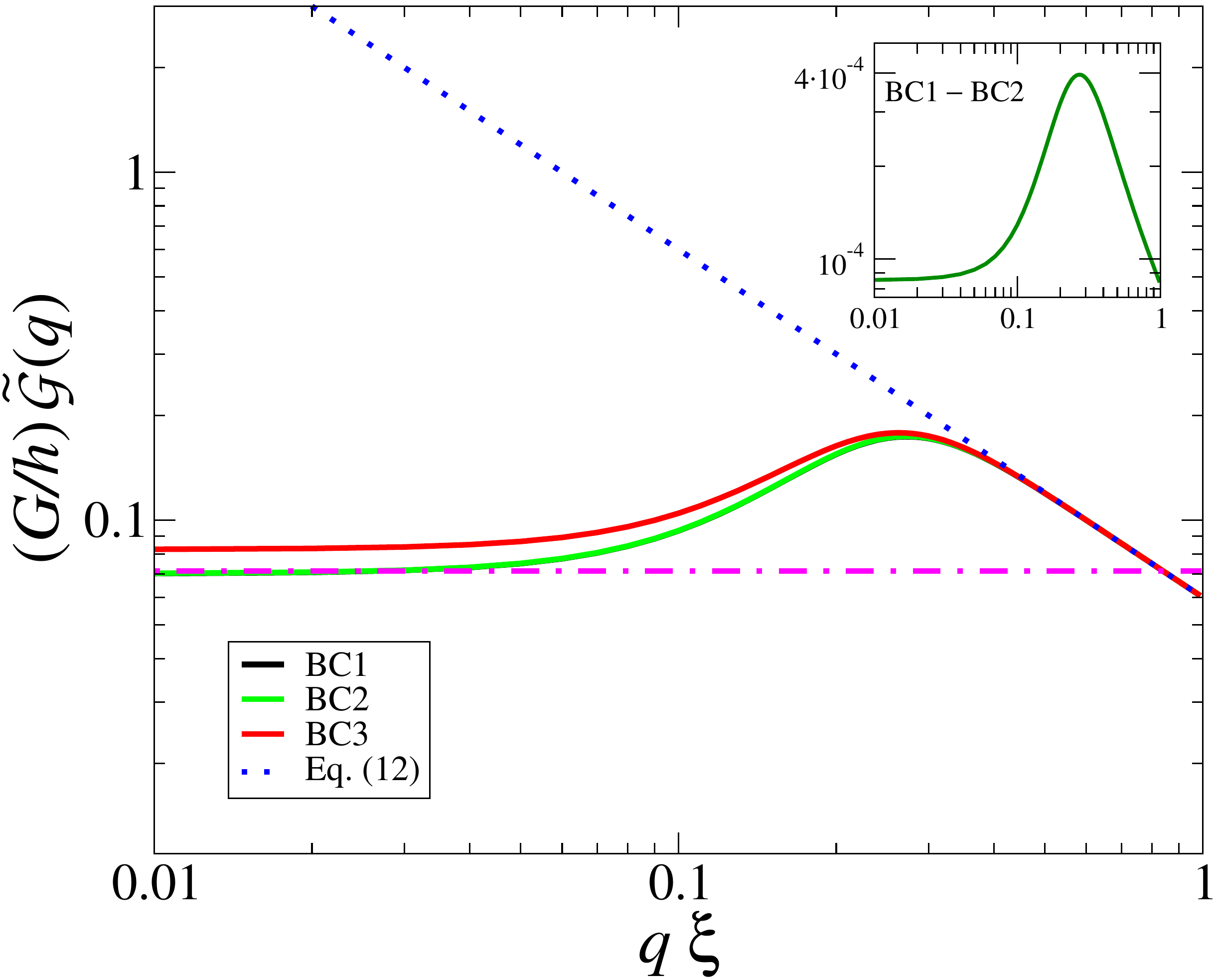}}
\caption{Perpendicular surface response, normalized by $h/G$, as a function of wavevector, normalized by $\xi^{-1}$. The normalized functions depend on the
  dimensionless parameters $h/\xi$, $\nu_0$, and $G/(i\omega\eta_0)$. We have used  $h/\xi=10$ and $\nu_0=0.4$. For $G/(i\omega\eta_0)$ we have taken large values ($> 10^3$), where the normalized function becomes independent of this parameter.  The solid lines show the results for the three sets of boundary conditions. The results for BC1 and BC2 are hardly distinguishable for this value of $h/\xi$ (lower curve, overlapping black and green); the small difference is shown in the inset. The upper red curve shows the results for BC3. The two asymptotes given by Eqs.~(\ref{G_P_eq}) and (\ref{Winkler_Asymptote_eq}) are shown by the dotted and dash-dotted lines, respectively.}
\label{fig_Gzz_qSpace}
\end{figure}

In the third step we extract from the solution the perpendicular response of the network at the surface,
\begin{equation}
    \tilde{u}_z(\vecq) = \tcG(\vecq) F_z.
\label{tcGq}
\end{equation}
In the present axially symmetric problem the response depends only on $q\equiv|\vecq|$. Figure~\ref{fig_Gzz_qSpace} shows the resulting response functions for the three boundary conditions (solid curves). Once $\tcG$ is normalized by $h/G$, and $q$ is normalized by $\xi^{-1}$, the response function depends on the
dimensionless parameters $h/\xi$, $\nu_0$, and $G/(i\omega\eta_0)$. To simplify the dependence on frequency, we hereafter focus on the common case of a sufficiently rigid network such that, for all relevant frequencies, $G/(i\omega\eta_0)\gg 1$. In this limit the response is inversely proportional to $G$, and thus its dependence on frequency is dominated by the prefactor $1/G(\omega)$. (Away from this limit, the dependence on $G/(i\omega\eta_0)$ is more complicated and cannot be presented concisely; fortunately, the deviation from the limit is hardly relevant experimentally.)
%\footnote{An additional, %much weaker dependence on %$\omega$ is included in %the Poisson ratio %$\nu(\omega)$.} 
Hence, after normalization by $h/G$, the response function becomes independent of $G/(i\omega\eta_0)$. The response functions for the first two sets of boundary conditions, BC1 and BC2 [Eqs.~(\ref{bc1}) and (\ref{bc2})], 
are hardly distinguishable (inset). At small and large $q$, the function has the following asymptotes:\footnote{The large-$q$ asymptote given in Eq.~(\ref{G_P_eq}) holds if $h\ll [G/(\omega\eta_0)]^{1/2}\xi\equiv\ell$. In the opposite limit of thick or soft substrate, $h\gg\ell$, the asymptote is the same for $q^{-1}\ll\ell\ll h$, and changes to $(2Gq)^{-1}$ for $\ell\ll q^{-1}\ll h$. The latter coincides with the perpendicular response of the Boussinesq problem (semi-infinite substrate). More details on these regimes can be found in Ref.~\cite{BarHaim2020}. They are less relevant in the present work, which focuses on the $qh\ll 1$ limit.}
\begin{equation}
  \tcG(q) = 
  \left\{\begin{array}{ll}
  \text{const},\ \ \ \ \ \ &qh\ll 1,\\
  \frac{1-\nu_0}{G} 
  \,\frac{1}{q},&qh\gg 1.
  \end{array}\right.
\label{G_P_eq}
\end{equation}
These asymptotes are also shown in Fig.~\ref{fig_Gzz_qSpace}.
The constant asymptote at $qh\ll 1$ is of particular interest, as will be discussed shortly. 

The final step is to relate the response function with the Winkler constant. In the limit $qh\ll 1$, $\tcG$ describes the perpendicular surface response at distances much larger than the substrate's thickness. This is exactly the limit where the Winkler-foundation description should hold. Indeed, in this limit $\tcG(q)\sim\text{const}$, which is inverted into $\cG(\vecrho)\sim\delta(\vecrho)$, \ie a completely local response. Thus the Winkler constant is directly related to the constant asymptote of the response at small $q$,
\begin{equation}
    k = \left[\tcG(q=0)\right]^{-1}.
\label{reduction}
\end{equation}
Equation~(\ref{reduction}) provides a general scheme for reducing three-dimensional elasticity into the Winkler surface response.

\begin{figure}
\centerline{\includegraphics[width=0.485\linewidth]{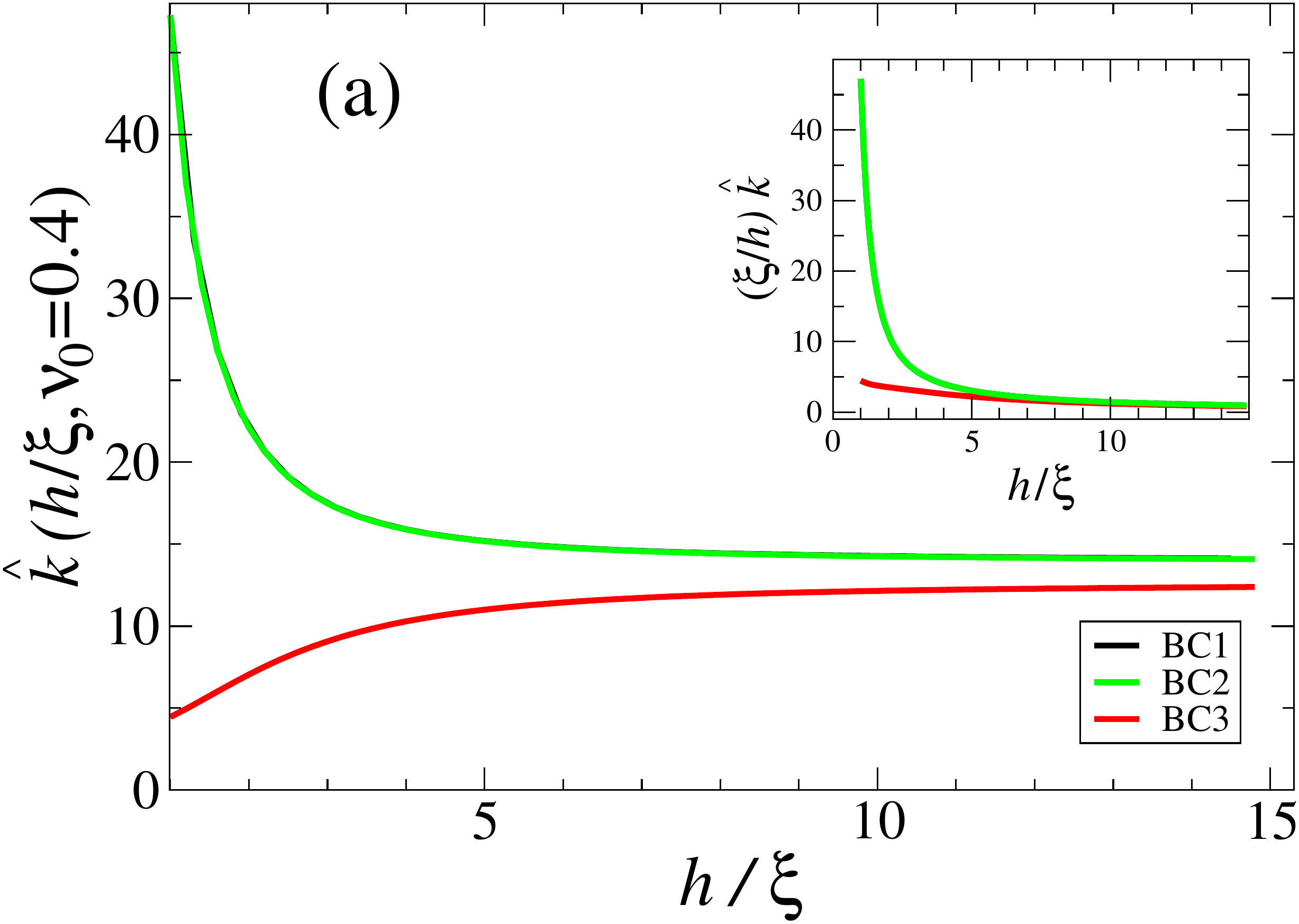}
\hspace{0.5cm} \includegraphics[width=0.46\linewidth]{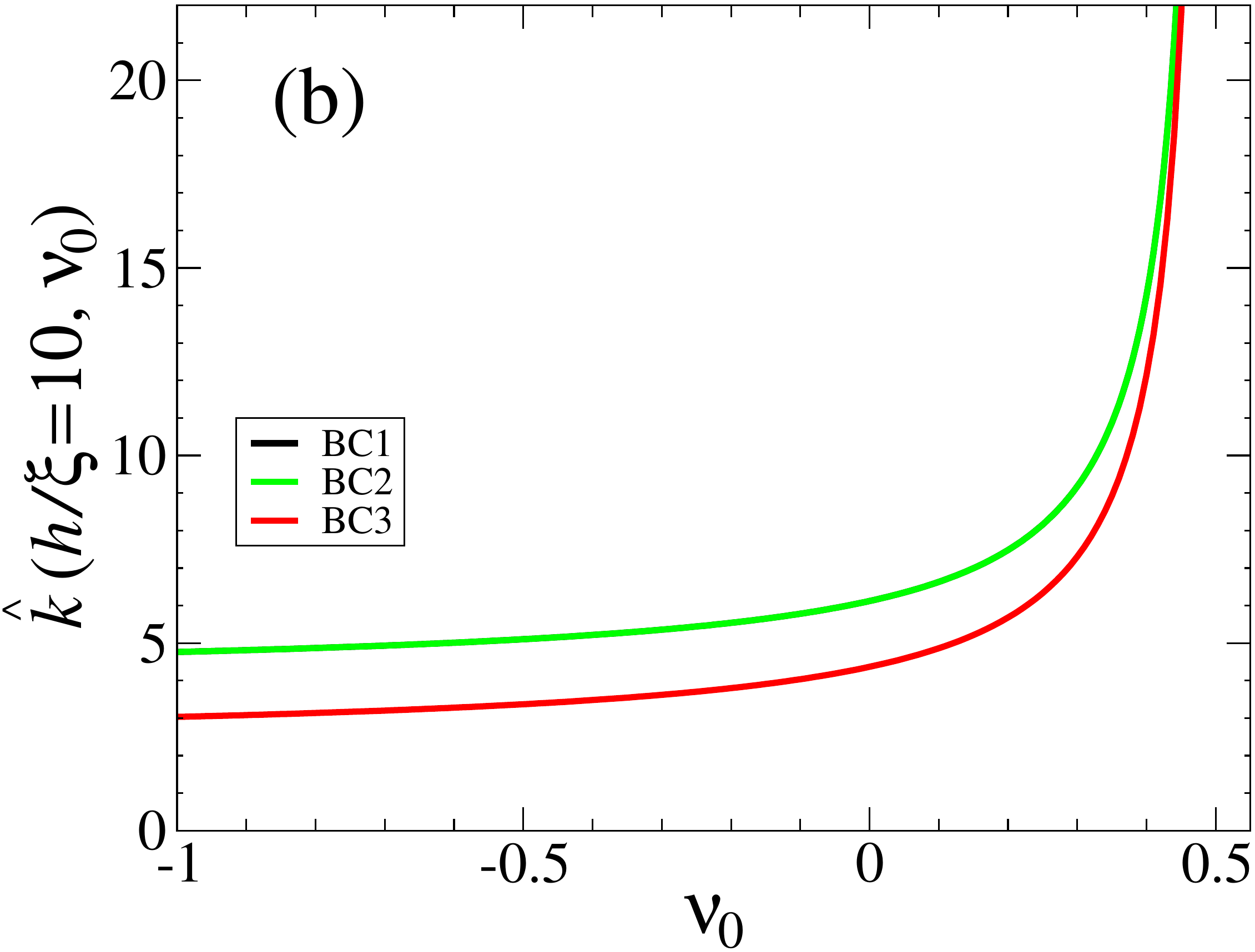}}
\caption{Dependencies of the Winkler constant, normalized by $G/h$, on film thickness and the network's Poisson ratio, for the three boundary conditions. In both panels the upper curve (overlapping black and green) shows the results for BC1 and BC2, and the lower red curve corresponds to BC3. Panel (a) shows the dependence on $h/\xi$ using $\nu_0=0.4$. The inset presents the same data multiplied by $\xi/h$, demonstrating that without the normalization by $G/h$ all boundary conditions yield decreasing functions. Panel (b) shows the dependence on $\nu_0$ using $h/\xi=10$. A large  value ($> 10^3$) has been taken for $G/(i\omega\eta_0)$, where the normalized Winkler constant becomes independent of this parameter.}
\label{fig_WinklerConst}
\end{figure}

Following the normalization and the large-stiffness assumption discussed above, we define a normalized Winkler constant which is a function of $h/\xi$ and $\nu_0$ alone,
\begin{equation}
    \label{k_hat_def_eq}
    \khat(h/\xi,\nu_0) \equiv (h/G) k.
\end{equation}
Figure~\ref{fig_WinklerConst} shows the dependencies of $\khat$ on $h/\xi$ for a given $\nu_0=0.4$ [panel (a)], and on $\nu_0$ for a given $h/\xi=10$ [panel (b)], for the three sets of boundary conditions. Once again, the differences between the results for BC1 and BC2 are insignificant. The behavior for BC3, Eq.~(\ref{bc3}), is markedly different, showing even an opposite trend as a function of $h$. Note, however, that after division by $h$ to remove the normalization, all curves decrease with $h$; see inset of panel (a). In addition, the asymptotic values of $\khat$ for large $h$ differ between BC1,2 and BC3 [panel (a)]. All results diverge for an incompressible substrate, $\nu_0\rightarrow 1/2$ [panel (b)].

Next, we specialize to the $qh\ll 1$ asymptote of $\tcG$ in the thick-substrate limit, $h/\xi\gg 1$. Asymptotic analysis yields 
\begin{equation}
\label{Winkler_Asymptote_eq}
    \text{BC1, BC2}:\ \ \ \ \khat(h/\xi\rightarrow\infty,\nu_0) =
    \frac{2(3-4\nu_0)}{1-2\nu_0}.
\end{equation}
For BC3 we have not been able to derive such a simple asymptote; as seen in Fig.~\ref{fig_WinklerConst}(a), the asymptotic value for BC3 differs from Eq.~(\ref{Winkler_Asymptote_eq}). 

The limit $h/\xi\gg 1$ holds also when $\xi\rightarrow 0$, \ie when the medium is taken as structureless. Thus the result from the two-fluid model should converge in this limit to the classical result for a structureless medium. Equation~(\ref{Winkler_Asymptote_eq}) coincides with Eq.~(\ref{Vlasov}), however, only in the sense that both diverge for $\nu=\nu_0=1/2$. This is because our composite medium has been taken as overall incompressible.\footnote{Similarly, the results of the two-fluid Boussinesq problem coincide in the limit $\xi\rightarrow 0$ with the classical one only for $\nu_0=1/2$ \cite{BarHaim2020}.} We have verified that, once the viscous component is removed from the model, Eq.~(\ref{Vlasov}) is recovered exactly.

\begin{figure}
\centerline{\includegraphics[width=0.48\linewidth]{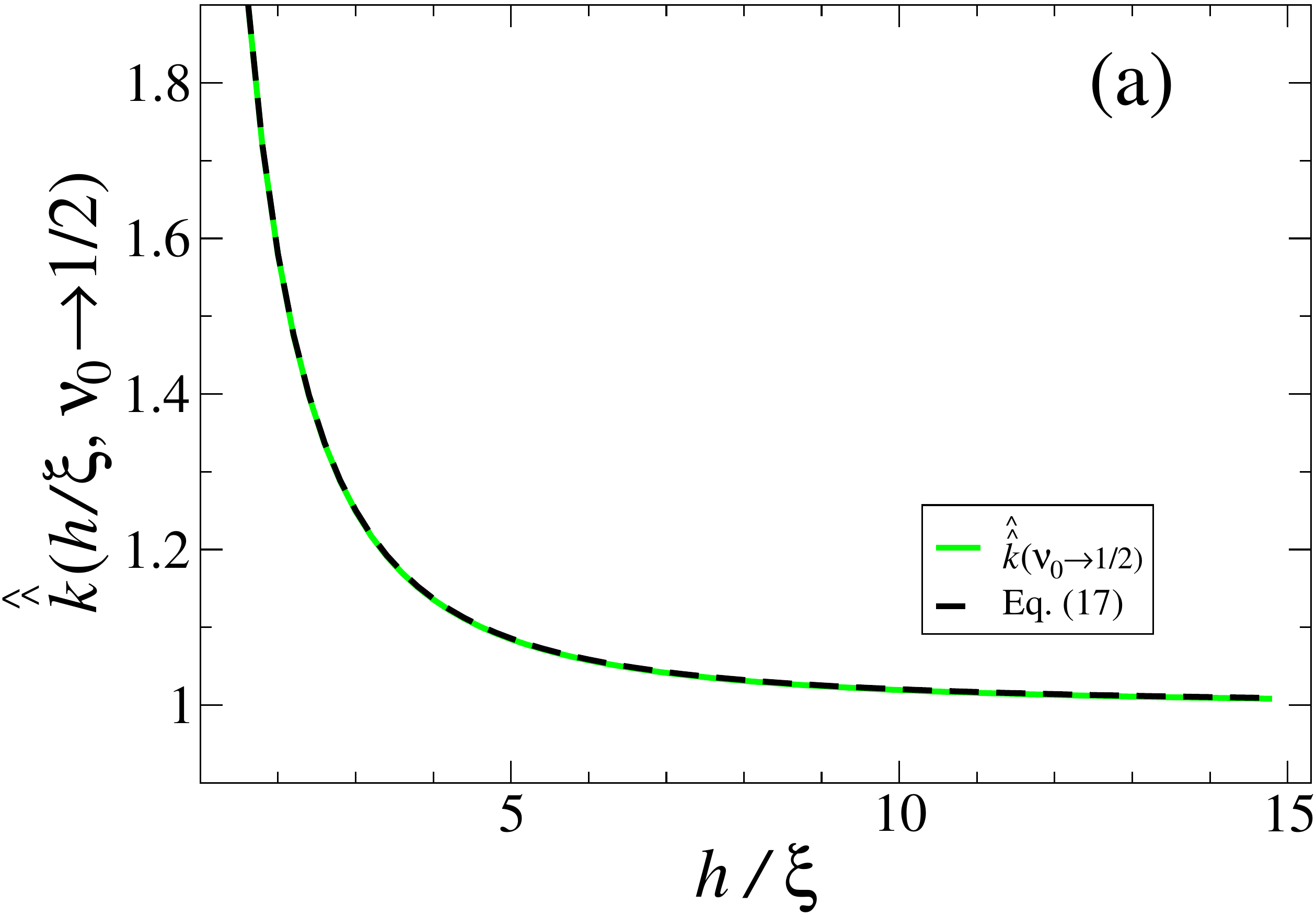}
\hspace{0.3cm} \includegraphics[width=0.48\linewidth]{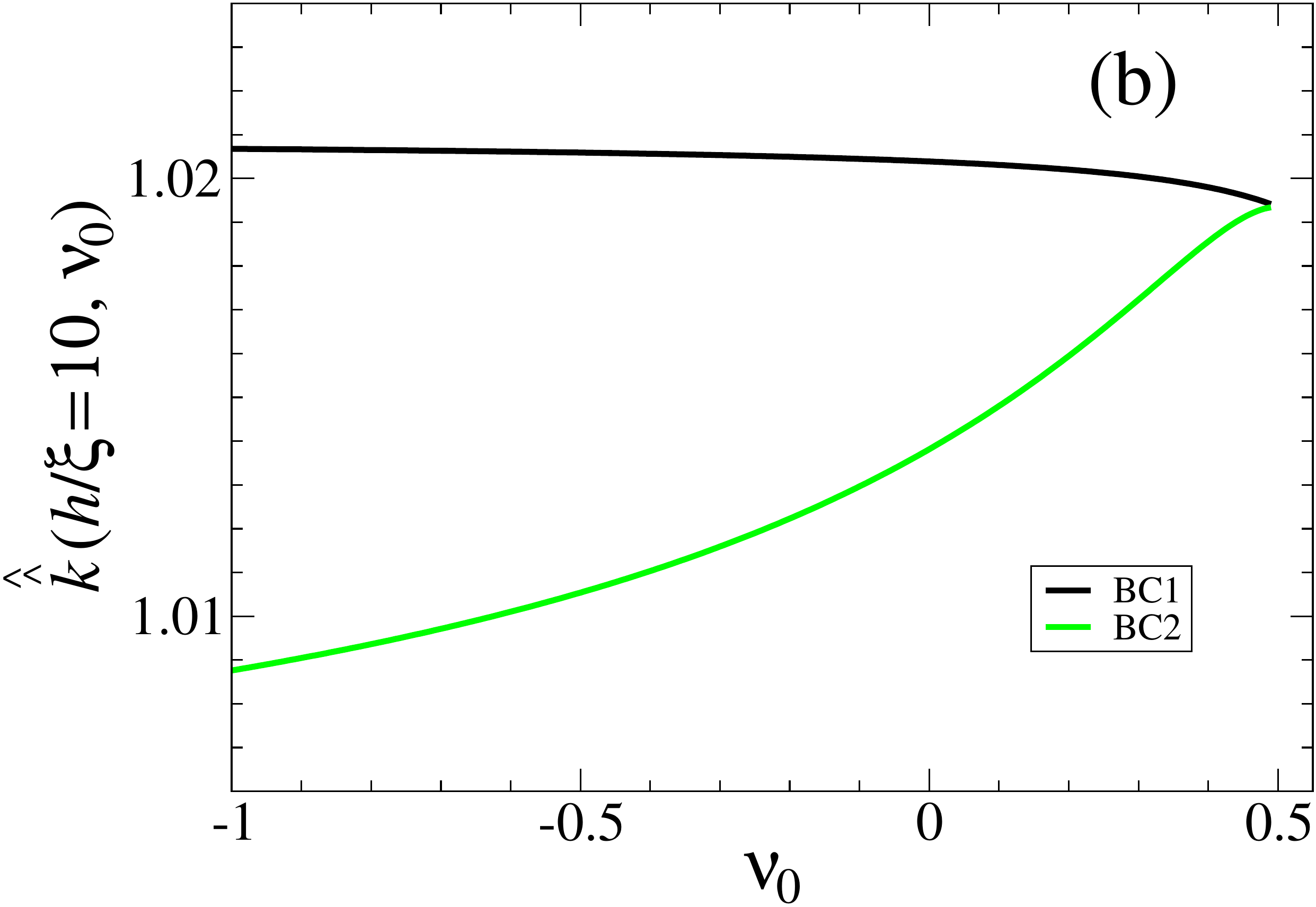}}
\caption{Deviations of the Winkler constant from its structureless (large-thickness) limit. The  panels show (a) the dependence on the normalized thickness $h/\xi$ for $\nu_0=1/2$, and (b) the dependence on the Poisson ratio $\nu_0$ for $h/\xi=10$, once the $h\rightarrow\infty$ value [Eq.~(\ref{Winkler_Asymptote_eq})] has been factored out. In (a) the two boundary conditions BC1 and BC2 give identical results [Eq.~(\ref{khat_h_eq})]. In (b) the deviations in the two cases (upper black curve for BC1; lower green curve for BC2) differ qualitatively. The dominant dependence, however, which has been factored out in this figure, is identical for the two boundary conditions. 
%For $G/(i\omega\eta_0)$ we have taken large %values ($> 10^3$), where the normalized %constant becomes independent of this %parameter.
}
\label{fig_FullyNormalizedWinklerConstBC1BC3}
\end{figure}

We are mainly interested in the effect of the substrate's structure (\ie $\xi$) on the Winkler constant. To examine the deviations of the results from the structureless limit of Eq.~(\ref{Winkler_Asymptote_eq}), we define the ratio,
\begin{equation}
    \hat{\khat}(h/\xi,\nu_0) \equiv \khat(h/\xi,\nu_0)/\khat(\infty,\nu_0).
\end{equation}
In particular, the divergence with $\nu_0\rightarrow 1/2$ has been scaled out. In this limit we obtain the closed-form expression,
\begin{equation}
    \text{BC1, BC2:}\ \ \ \ \hat{\khat}(h/\xi,1/2) = \frac{(h/\xi)^2(1+e^{-2h/\xi})}{(h/\xi)^2-2+4e^{-h/\xi}+[(h/\xi)^2-2]
    e^{-2h/\xi}}.
\label{khat_h_eq}
\end{equation}
In the limit $h\rightarrow\infty$ this expression reduces to $1$ as required. The leading-order correction in small $\xi/h$ is
\begin{equation}
    \text{BC1, BC2, $\xi/h\ll 1$:}\ \ \ \
    \hat{\khat}(h/\xi,1/2) \simeq
    1 + 2\xi^2/h^2.
\label{khat_large_h}
\end{equation}
Figure~\ref{fig_FullyNormalizedWinklerConstBC1BC3}(a) shows the thickness dependence of $\hat{\khat}$ for $\nu_0=1/2$ according to Eq.~(\ref{khat_h_eq}). Appreciable deviation (at least 1 percent) from the structureless limit is found for $h\lesssim  10\xi$. Finally, in Fig.~\ref{fig_FullyNormalizedWinklerConstBC1BC3}(b) we examine how the dependence on the Poisson ratio deviates from its large-$h$ asymptote, Eq.~(\ref{Winkler_Asymptote_eq}), for $h=10\xi$. The deviation is small, about 2\%, for all reasonable values of $\nu_0$. For $\nu_0\rightarrow 1/2$ the results for BC1 and BC2 converge, as Eq.~(\ref{khat_h_eq}) holds for both boundary conditions. For lower values of $\nu_0$ the behavior slightly differs between BC1 and BC2. Recall, however, that the dominant dependence on $\nu_0$, Eq.~(\ref{Winkler_Asymptote_eq}), has been factored out in Fig.~\ref{fig_FullyNormalizedWinklerConstBC1BC3}(b). This dominant behavior is identical for BC1 and BC2.

\section{Discussion}
%-------------------
\label{sec_discussion}

Out of the various results obtained above, the most useful for experiments may be the prediction for the Winkler constant in the limit of thick (but finite), close to incompressible, structured substrate. Collecting Eqs. (\ref{k_hat_def_eq}), (\ref{Winkler_Asymptote_eq}), and (\ref{khat_large_h}), we get the relation
\begin{equation}
    \label{eq_final}
    h\gg\xi,\ \nu_0\lesssim 1/2:\ \ \ 
    k = \frac{G(\omega)}{h}
    \frac{2(3-4\nu_0)}{1-2\nu_0}\, \left(1 + \frac{2\xi^2}{h^2} \right),
\end{equation}
which is valid for both boundary conditions BC1 [Eq.~(\ref{bc1})] and BC2 [Eq.~(\ref{bc2})]. The more elaborate Eq.~(\ref{khat_h_eq}) readily extends this relation to films whose thickness is not much larger than their correlation length $h\gtrsim\xi$. These relations can be used to extract the substrate's viscoelastic shear modulus $G(\omega)$, correlation length $\xi$, and the network's Poisson ratio $\nu_0$ (an illusive property) from various surface measurements, \eg the wrinkling of a thin rigid sheet supported on the substrate.

The relation between Eq.~(\ref{eq_final}) and the simple viscoelastic result, Eq.~(\ref{eq_vlasov}), is subtle, since $\nu$ (the Poisson ratio of the bulk material) and $\nu_0$ (the Poisson ratio of the bare, solvent-free network) are different parameters. Only when both $\nu$ and $\nu_0$ are equal to $1/2$, do both the structureless and two-fluid models describe an overall incompressible material, and the two results coincide\,---\,they diverge. (The same was found in Ref.~\cite{BarHaim2020} for two non-divergent results.) The practical conclusion is the following. If the film is appreciably far from the incompressible limit, and its thickness is much larger than its correlation length, then Eq.~(\ref{eq_vlasov}) can be used. If it is essentially incompressible, or $h$ is not much larger than $\xi$, Eq.~(\ref{eq_final}) should be used. The important novelty to bear in mind, however, is the dependence of Eq.~(\ref{eq_final}) on the correlation length $\xi$.

We have treated three sets of boundary conditions, reflecting different couplings between the constituents of the structured material at its surface, and between them and the forcing object. In BC1, Eq.~(\ref{bc1}), the material's elastic and viscous components are strongly coupled at the surface, and the force is exerted on both. This fits a material with large surface tension (strong attraction between the constituents), where a forced bead, for instance, is in physical contact with the surface. In BC2, Eq.~(\ref{bc1}), the two components are strongly coupled as well, but the force is exerted on the elastic component alone. This corresponds, \eg to an experiment where the forced particle is chemically attached to the polymer network. The Winkler constants for these two cases, however, have been found to be almost indistinguishable, regardless of the values of $h$ or $\nu_0$. The physical reason may lie in the little freedom left for the material's two components when they are strongly coupled at the surface and are both stationary at the opposite-side rigid boundary. Thus forcing either one of them or both hardly matters. Indeed, using BC3, Eq.~(\ref{bc3}), where the two components are weakly coupled at the surface, yields different results; see, \eg Fig.~\ref{fig_WinklerConst}. Still, for $h\gtrsim 10\xi$ and $\nu_0\gtrsim 0.4$, the difference amounts to a few percent.

We have not considered explicitly the normal force arising from the substrate's surface tension. The reason is that it should have no effect on the Winkler constant, as it is associated with deformation gradients beyond the purely local Winkler response. Indeed, while a finite surface tension $\gamma$ introduces into the normal surface response a term $\sim\gamma q^2$, the Winkler constant is obtained from the limit $qh\rightarrow 0$ (cf.\ Fig.~\ref{fig_Gzz_qSpace}).

Focusing on the effect of the substrate's viscoelasticity and structure, we have not addressed the known divergence in the incompressible limit ($\nu_0\rightarrow 1/2$), as mentioned in the Introduction. A possible resolution of this problem may probably be achieved by going to the next (quadratic) order in $qh$ and performing the appropriate asymptotic analysis. This should extend the results of Ref.~\cite{Chandler2020} to the case of a structured substrate.

Another interesting extension of the present theory would be to derive the nonlinear correction to the Winkler model \cite{Brau2011} for a structured material. This will allow a study of richer phenomena, such as elaborate pattern formation in thin sheets supported on such substrates.

\acknowledgments

We thank Tom Witten for a helpful discussion and his insight
concerning the divegence of $k$. This research has been supported by
the Israel Science Foundation (Grant No.\ 986/18).

\bibliography{Winkler_bibl}

\end{document}